\documentstyle[preprint,aps,eqsecnum,floats]{revtex}
\begin{document}

%\date{\today}

\draft
\preprint{\vbox{Submitted to Physical Review D
                \hfill DOE/ER/40762--041\\
                \null\hfill U. of MD PP \#94--177}}

\title{Exclusive Production of Higgs Bosons in Hadron Colliders
       \thanks{Work supported by the Department of
               Energy, contract DE--FG02--93ER--40762.
              }
      }

\author{Hung Jung Lu and Joseph Milana}

\address{Department of Physics, University of Maryland \\
         College Park, Maryland 20742
        }
\date{\today}
\maketitle

\begin{abstract}
We study the exclusive, double--diffractive production of the
Standard Model Higgs particle in hadronic collisions at
LHC and FNAL (upgraded) energies.  Such a mechanism would
provide an exceptionally clean signal for experimental detection
in which the usual penalty for triggering on the rare decays
of the Higgs could be avoided.  In addition, because of the
color singlet nature of the hard interaction, factorization
is expected to be preserved, allowing the cross--section
to be related to similar hard--diffractive events at HERA.
Starting from a Fock state expansion in perturbative QCD,
we obtain an estimate for the cross section in terms of
the gluon structure functions squared of the colliding hadrons.
Unfortunately, our estimates yield a production rate well
below what is likely to be experimentally feasible.
\end{abstract}

\vspace{0.25in}


\section{Introduction}
The field of hard diffractive scattering has acquired a
solid experimental basis with the report last year
\cite{UA8} of the UA8 data observing jet events
in conjunction with an elastically scattered hadron
at the S$p \bar p$S collider at CERN.  Similar findings
\cite{RapGapZEUS} by ZEUS of large rapidity gap, jet events
at HERA are also suggestive of such diffractive events,
although the existence of a final state proton still needs to be
confirmed.  One interesting application of such diffractive events
would be in the search for fundamental new physics, Higgs bosons.
Such possibilities have indeed been discussed
before\cite{SchaeferNachtmannSchoepf,MuellerSchramm,BialasLandshoff,BJ},
although
mostly in the context of inclusive, diffractive processes:
\begin{equation}
p + p \rightarrow H + p + p + X.
\end{equation}
Here we are interested on the other hand with the truly exclusive,
double--diffractive production:
\begin{equation}
p + p \rightarrow H + p + p.\label{exclusive}
\end{equation}

This interest is two fold.  First, because the
 experimental signal for this process would be
exceptionally clean with the final state hadrons well
separated in rapidity from the Higgs decay fragments, it should be possible
 to trigger on the dominant decays of the Higgs particle,
($b\bar b$ jet-pairs for $m_H < .8$ TeV)
rather than on a rare decay mode as is necessary in the case of inclusive Higgs
production\cite{HiggsHunter,ColliderPhys} due to the otherwise unmanageable
backgrounds.  To the extent that
(\ref{exclusive}) does occur, one would expect, due to the narrow width of a
Standard--Model Higgs (for instance, $\Gamma(H \rightarrow all)
\approx \Gamma(h \rightarrow b \bar b) = 2.5$ MeV for $m_H = 100$ GeV),
 that it would stand out rather prominently
above the corresponding background of double--diffractive $b\bar b$ jet-pairs.
The use of the silicon--vertex technology by which CDF\cite{CDFtop}
is now exploiting
to tag top decays into bottom quarks should particularly help suppress the
background, eliminating the analogous jet events of light parton
($gg, u \bar u, d \bar d$, etc.) pairs.

Secondly, it is our expectation that events such as (\ref{exclusive}) (and
their background) should obey factorization.  In comparison, estimates
for inclusive production of the Higgs
\cite{SchaeferNachtmannSchoepf,MuellerSchramm,BialasLandshoff}
are likely to be less reliable due to the expected breakdown
of factorization.  These
expectations are based upon the observation of Collins, Frankfurt, and
Strikman\cite{CollinsFrankfurtStrikman} (see also Ref. \cite{BereraSoper} in
this
regard) that in the case that a hard
scattering event produces a color octet state, it is impossible to disentangle
the soft and hard physics processes involved in diffractive events.  In more
technical terms, the limitation in the final state to an elastically scattered
hadron prevents the sum over all possible cuts of the hard scattering
graphs necessary to eliminate soft gluon exchanges in the
proofs of factorization\cite{factproofs} for inclusive processes.
These authors then predict that in the case of ``coherent diffractive events'',
where all the momentum
lost by the proton appears in the hard scattering event, factorization should
be
violated at the leading twist level, showing up most dramatically in dijet
diffractive
events at HERA in which such ``coherent'' events
would be higher--twist relative to the ``incoherent'' diffractive events due to
the
color--singlet nature of the hard interaction.  While agreeing with these
general
arguments we pointed out in Ref. \cite{LuMilana} that some of these predictions
may nevertheless not be so obvious due to the potential complications in
these rates of an additional expansion in $1/\xi$,
the energy loss of the diffracted proton.  (Depending upon the low energy
behavior of the soft physics, the higher--twist suppression might be
compensated
at least in part, by differing powers in $1/\xi$ between the
``coherent'' vs. ``incoherent'' mechanisms.)  Irrespective though of these
complications,
in the present case we do not believe the considerations of
\cite{CollinsFrankfurtStrikman}
should apply as the hard scattering event now involves a
color singlet process, and hence, although in itself higher--twist, additional
soft gluon exchanges should be suppressed.   Indeed we believe all such
exclusive
double--diffracted, jet events should be added to the otherwise exhaustive list
of
experimental processes in Ref. \cite{CollinsHustonetc.} for exploring the
nature of hard
diffractive scattering.

In Ref. \cite{MuellerSchramm} an estimate of the cross
section for Higgs production via exclusive double diffraction in a Pomeron
based
model was obtained,
with a value $\sigma \sim 10^{-6}$ pb for a Higgs mass of $150$ GeV.
Although this is a small cross section, nonetheless it
might be accessible for measurement given the uncertainty
in the theoretical prediction or an improvement in
experimental luminosity.  Indeed at the planned luminosities of the LHC
(${\cal{L}} = 1.7 \times
10^{34} \, \rm(cm^{-2}-sec^{-1})$ \cite{PartData}), such a rate could
correspond to a half--dozen or so events per year at these Higgs masses,
depending
upon the overall coefficient assignable to this order of magnitude estimate.
This in itself is competitive with the present status of top quark
events being searched for at FNAL\cite{CDFtop,D0top}.  On the other hand,
 due to the steep dependence with
Higgs mass of the rate for (\ref{exclusive}), significantly more abundant
yields
 (hundreds) would be anticipated if the mass of the
 Higgs was at the lower end of the presently allowable values (
 $m_H > 64$ GeV\cite{LEPbound}) assuming the estimates of \cite{MuellerSchramm}
were correct.  Given the
cleanness of the signal for the Higgs
boson, it is important to further
explore this mechanism.  In particular, it is highly
desirable to have an alternative and independent estimate
of the production cross section.

Note that we will not consider here Higgs production through photon--photon
fusion.  This mechanism, considered elsewhere\cite{photonphoton}
(although predominantly in the
context of nuclear--nuclear collisions and for collider parameters more
ambitious than being presently projected\cite{PartData}) also produces the
desired
double--diffracted signature.  However it involves coherent radiation
of photons off of the entire hadron and hence are not amenable to a
partonic type analysis to be employed here.   All results should therefore
be understood to apply to these latter
(``Pomeron--Pomeron'' fusion) type of mechanisms.

In a previous paper \cite{LuMilana}, we employed a Fock state
expansion in perturbative QCD to compute hard--diffractive events in
electron--proton collisions.
We showed that in this approach, and consistent with other estimates
in related processes\cite{Ryskin,Brodskyetal}, the unknown
soft matrix element that entered the amplitude for these events at HERA
can be related to the gluon
structure function in the low-$x$ region.
In this paper we apply this method to
the exclusive production of Higgs bosons at hadron--hadron colliders.
In addition to the generic triangle loop
diagram considered in Refs.
\cite{SchaeferNachtmannSchoepf,MuellerSchramm,BialasLandshoff,BJ},
we have included
in our analysis the creation of the Higgs particle through box and
pentagon top--quark loops.  Indeed in our approach these contribute
roughly equally in
the amplitude for the process being considered and to our best knowledge,
have not been considered previously.

Unfortunately our estimates yield that the
cross section at either upgraded Fermilab (FNAL$^*$) energies
or at the LHC is particularly small and
hence, despite the potential cleanness of the signal,
we do not expect the double--diffractive mechanism to
be useful at either laboratory for discovering the Higgs boson.

\section{Effective Higgs-Gluon Interaction}
As we will argue in the next section, the dominant
contribution to the exclusive production of Higgs
comes from processes involving gluons in the
initial and final state. Hence we will need to know
the interaction between gluons and Higgs. In particular,
we will need the effective interaction vertices of
two, three and four gluons with a Higgs particle.

The Higgs particle does not couple directly to gluons.
However, it can do so through a quark loop. Since the
Yukawa coupling of the Higgs particle is proportional
to the mass of the coupled fermion, we need only to
consider the top-quark loop. The effects of lighter
quarks ($u, d, s, c, b$) are negligible due to the
substantially heavier mass of the top quark.
We will consider the regime $m_H \le m_t$, where
$m_H$ is the Higgs mass and $m_t$ the top mass.
As we will see later, the Higgs production cross
section decreases with increasing Higgs mass.
Therefore, the $m_H \le m_t$ regime is where one
can expect a higher cross section. The restriction
to the case of lighter Higgs masses has the technical
benefit that it significantly facilitates the computation
of the Feynman integrals since we can compute
all integrals to leading order in $m_H/m_t$.
The expansion in $m_H/m_t$ is better in fact than one might naively expect,
as the relevant parameter is the ratio of the two masses
multiplied by a typical value of one or more Feynman parameters entering
the particular loop's evaluation, so that even at $m_H/m_t \approx 1$,
corrections  are small (see Ref. \cite{DawsonKauffman} for
explicit calculations of the higher corrections to this approximation in the
case of inclusive Higgs production).  Indeed, the more Feynman parameters
(heavy quark propagators) the better the approximation.

In this limit, the heavy top-quark loop integrates
out and the Higgs particle becomes effectively
coupled to the gluons. In Fig. \ref{fig1}
we present the effective interaction vertices $ggH, gggH, ggggH$
between the Higgs particle and gluons. Notice in
particular the momentum flow convention for the
$ggH$ and $gggH$ vertices. In the figure,
$a,b,c,d$ represent the color indices and
$\lambda,\mu,\nu,\sigma$ the Lorentz indices
of the external gluons.

The following are the analytical results for the
interaction vertices, where $g_{\rm w}$ and $g_{\rm s}$
are the weak $SU(2)$ and the strong $SU(3)$
coupling constant, and $m_W$ the mass
of the $W$ gauge boson.

\begin{eqnarray}
i G_{\lambda\mu}^{ab} (k_1, k_2)
&=&
\frac{i}{24\pi^2}
\frac{g_{\rm w} g_{\rm s}^2}{m_W}
\delta^{ab}
( g_{\lambda\mu} k_1 k_2 - k_{2 \lambda} k_{1 \mu} ) ,
\label{EqGGH}
\\
i G_{\lambda\mu\nu}^{abc} (k_1, k_2, k_3)
&=&
\frac{1}{24\pi^2}
\frac{g_{\rm w} g_{\rm s}^3}{m_W}
f^{abc}
\left[
 g_{\lambda\mu} (k_1-k_2)_\nu
+g_{\mu\nu}     (k_2-k_3)_\lambda
+g_{\nu\lambda} (k_3-k_1)_\mu
\right] ,
\label{EqGGGH}
\\
i G_{\lambda\mu\nu\sigma}^{abcd} (k_1, k_2, k_3, k_4)
&=&
\frac{i}{24\pi^2}
\frac{g_{\rm w} g_{\rm s}^4}{m_W}
\left[
 f^{abe}f^{cde}
 ( g_{\lambda\nu}g_{\mu\sigma} - g_{\lambda\sigma}g_{\mu\nu} )
\right.
\nonumber
\\
& & \hspace{0.64in}
+ f^{ace}f^{bde}
 ( g_{\lambda\mu}g_{\nu\sigma} - g_{\lambda\sigma}g_{\mu\nu} )
\nonumber
\\
& & \hspace{0.62in}
\left.
+ f^{ade}f^{bce}
 ( g_{\lambda\mu}g_{\nu\sigma} - g_{\lambda\nu}g_{\mu\sigma})
\right] .
\label{EqGGGGH}
\end{eqnarray}
Notice the $ggH$ vertex satisfy the conditions
\begin{equation}
  k_1^\lambda \ i G_{\lambda\mu}^{ab}(k_1, k_2)
= 0
= i G_{\lambda\mu}^{ab}(k_1, k_2) \ k_2^\mu .
\end{equation}
The expression for the triangle diagram Eq. (\ref{EqGGH})
has been known for sometime \cite{triangle}.
Note the similarity
between the $gggH$ and $ggggH$ vertices
with the conventional form of the pure-QCD
three- and four-gluon vertices. Observe,
however, that the three momenta $k_1, k_2, k_3$ in
the $gggH$ vertex are independent variables,
as opposed to the case of the pure-QCD three-gluon
vertex. Observe also that the expression for the
$ggggH$ vertex does not contain an explicit dependence
on the external momenta, just like in the case of the pure-QCD
four-gluon vertex. Finally, notice that the effective
gluon-Higgs vertices do not depend on the value of
the top-quark mass (this is valid only for  the $m_H \le m_t$
regime.)

\section{Scattering Amplitude}
The leading diagram for the exclusive production of Higgs
bosons is depicted in
Fig. \ref{fig2}. This figure can be interpreted as
the production of a Higgs particle through pomeron-pomeron
fusion, where the two gluon lines belonging to the same
proton effectively represent the pomeron interaction.
In principle the partons coming out of the protons could
also be quarks, in which case Higgs production would
occur through two possible
mechanisms: 1) fusion of virtual $Z^0 Z^0$ pairs radiated off of
diffracted quarks, or 2) $q \bar q$ fusion with
 an additional hard gluon pair creating the same flavor quarks back
into the colliding hadrons.  In our previous work
\cite{LuMilana} on exclusive dijet diffractive production at HERA,
such quark initiated contributions were suppressed due to an anticipated
less singular behavior
in the small--$x$ region of the soft matrix elements entering
the hard amplitude, occurring because of a partial cancellation between quark
and antiquark graphs.   This thus lead us to
 approximate the soft matrix elements by the valence quark
distribution functions.  In the present case such a cancellation only occurs
in the case of $Z^0$ particle fusion, for which we estimate a suppression
in the amplitude of
the order of $10^{-2}$ arising (as we will see below)  from the
ratio of valence quark to gluon structure functions squared.
In the case of $q \bar q$ fusion,
due to the explicit mass dependence of the coupling of quarks
to the Higgs, these graphs will also be suppressed either directly through the
vanishingly small couplings ($m_u \approx m_d \approx m_s \approx m_c \approx
0$)
or because of the highly suppressed probabilities of finding
bottom quarks (relative to gluons) in a proton.

Having now argued why only the
gluon initiated processes need be considered, we proceed with our evaluation
of Fig. \ref{fig2} using perturbative QCD methods.
Many of the steps in our approach
were already detailed in Ref. \cite{LuMilana}; we repeat them here for the sake
of completeness.
In the Fock component expansion \cite{BrodskyLepage,Mueller},
a proton can be expressed in terms of its parton (quark and gluon)
components as
\begin{equation}
\mid p \rangle = \sum_{n} \int [dx] [d^2k_{\perp}]
\frac{1}{\sqrt{x_1\cdot \cdot \cdot x_n}}
\psi_n(x_i,k_{i\perp}) \mid x_i,k_{i\perp} \rangle,
\label{EquationProtonFockExpansion}
\end{equation}
where $x_i$ are the longitudinal momentum fraction of the
proton carried by the $i$-th component parton and
$k_{i\perp}$ the corresponding transverse momentum.
We have suppressed color and spin indices here.
The individual parton states are normalized by
\begin{equation}
\langle p^{\prime}_i \mid p_j \rangle
=
16 \pi^3 x_i \delta(x^{\prime}_i - x_j) \delta^{(2)}
(k^{\prime}_{i\perp} - k_{j\perp})
\label{EquationPartonNormalization}
\end{equation}
and the measure is defined by
\begin{equation}
[dx] [d^2k_{\perp}] =
\left(
   \prod_{i=1}^{n} dx_i
   \frac{d^2k_{i\perp}}{16\pi^3}
\right)
\delta (1 - \sum_{j=1}^{n} x_j )
16\pi^3\delta^{(2)}( \sum_{j=1}^{n}k_{j\perp} ).
\label{EquationIntegrationMeasure}
\end{equation}
The wavefunctions obey the constraint
\begin{equation}
1 = \sum_{n} \int [dx] [d^2k_{\perp}] \mid \psi_n(x_i,k_{i\perp})
\mid^2,\label{normw}
\end{equation}
so that one obtains the canonical normalization for the proton that
\begin{equation}
\langle p \mid p^{\prime} \rangle = 16 \pi^3 E \delta^3
(p - p^{\prime}).\label{normp}
\end{equation}

Let $\xi_1$ be the fraction of momentum loss of the
first hadron and $\xi_2$ the corresponding fraction
of momentum loss of the second hadron.  We thus have for
Fig. \ref{fig2}
\begin{eqnarray}
p'_1 &=& (1-\xi_1) p_1, \\
p'_2 &=& (1-\xi_2) p_2.
\end{eqnarray}
In Fig. \ref{fig3} we show the Feynman diagrams for the
hard matrix element involving four external gluons
and one Higgs particle. Fig. \ref{fig3}a, \ref{fig3}b,
\ref{fig3}c represent typical Feynman diagrams for
the $s$-channel, $u$-channel and the four-gluon-vertex
exchange mechanisms. As discussed in the previous section,
the effective gluon-Higgs coupling vertices are induced
by a heavy top-quark loop. The Higgs line can be
attached to any gluon line, three-gluon vertex or
four-gluon vertex. Hence, we have 7 diagrams for the
$s$- or $u$-channel exchange mechanism, and 5 diagrams
for the four-gluon-vertex exchange. The effective
$ggH, gggH, ggggH$ vertices are given in
Eq. (\ref{EqGGH}), (\ref{EqGGGH}) and (\ref{EqGGGGH}).

At the parton level, we define $x_1$ to be the
fraction of momentum of hadron one, $p_1$, carried by the
first incoming gluon and $x'_1$ the fraction
of momentum of the diffracted hadron one, $p'_1$, carried by
the first outgoing gluon.
The momentum fractions $x_1$ and
$x'_1$ are related by
\begin{equation}
x_1 = \xi_1 + x'_1 (1-\xi_1) .
\label{EquationRelationXXPrime}
\end{equation}
Similarly, we define $y_1$ and $y'_1$
to be the corresponding initial and final
momentum fractions of the gluon from the
second hadron. These momentum fractions
are related by
\begin{equation}
y_1 = \xi_2 + y'_1 (1-\xi_2) .
\label{EquationRelationYYPrime}
\end{equation}

The kinematics of hard diffraction considerably
simplifies the calculation.  In particular,
the gluon polarization vectors are orthogonal to
the four--momenta of all the incoming and outgoing
gluons, hence a large number of dot products vanish.

Defining $\tilde x_1=(1-\xi_1) x'_1$ and
$\tilde y_1=(1-\xi_2) y'_1$,
averaging over initial helicity and color, and
considering only the case of final gluons that
conserve the initial color and helicity, the
scattering amplitude from the $s$-channel contribution
(or $u$-channel contribution, which turns out to be
identical) is given by
\begin{equation}
i {\cal M}^{(s,u)} =
 \frac{i}{16}
 \frac{g_{\rm w} \alpha_{\rm s}^2}{m_W}
\left[
-2
+\frac{x_1}{\tilde x_1}
+\frac{\tilde x_1}{x_1}
+\frac{y_1}{\tilde y_1}
+\frac{\tilde y_1}{y_1}
-\frac{2 x_1 y_1}{\tilde x_1 \tilde y_1}
-\frac{2 \tilde x_1 \tilde y_1}{x_1 y_1}
\right] ,
\end{equation}
where $g_{\rm w}$ is the $SU(2)$ weak coupling constant and
is related to the Fermi constant by
$G_{\rm F}/\sqrt{2}=g_{\rm w}^2/8 m_W^2$
(to tree level), with $m_W$ the mass of the
$W$ boson. And $\alpha_{\rm s}$ is the strong
coupling constant. Analogously, the total contribution
from the four-gluon-vertex diagrams, averaged over initial
helicities and colors, is given by
\begin{equation}
i {\cal M}^{(4g)} =
 \frac{i}{8}
 \frac{g_{\rm w} \alpha_{\rm s}^2}{m_W}
\left[
 2
-\frac{x_1}{\tilde x_1}
-\frac{\tilde x_1}{x_1}
-\frac{y_1}{\tilde y_1}
-\frac{\tilde y_1}{y_1}
\right] .
\end{equation}
The total hard amplitude is simply the sum of the $s$-channel,
$u$-channel and the four-gluon-vertex contributions:
\begin{equation}
i {\cal M}^{\rm hard}
=
i {\cal M}^{(s)} + i {\cal M}^{(u)} + i {\cal M}^{(4g)}
=
-\frac{i}{4}
 \frac{g_{\rm w} \alpha_{\rm s}^2}{m_W}
\left[
 \frac{x_1 y_1}{\tilde x_1 \tilde y_1}
+\frac{\tilde x_1 \tilde y_1}{x_1 y_1}
\right] .
\end{equation}

To obtain the scattering amplitude of $p+p \to p+p+H$
at the hadronic level, we must sandwich the hard
amplitude between the bras and kets of the incoming
and outgoing hadrons. Using Eq.
(\ref{EquationProtonFockExpansion}--\ref{EquationIntegrationMeasure}),
\begin{eqnarray}
i {\cal M}
&=&
\langle
       p'_1 p'_2
       \mid
           i {\cal M}^{\rm hard}
       \mid
       p_1 p_2
\rangle
\nonumber
\\
&=&
\sum_n (1 - \xi_1)^{(n-1)/2} \int \frac{[dx']
[d^2k'_{\perp}]}{\sqrt{x'_1 x_1}}
\psi^{*}_n(x',k'_{\perp}) \psi_n(x, k'_{\perp})
\nonumber
\\
& &
\sum_m (1 - \xi_2)^{(m-1)/2} \int \frac{[dy']
[d^2q'_{\perp}]}{\sqrt{y'_1 y_1}}
\psi^{*}_m(y',q'_{\perp}) \psi_m(y, q'_{\perp})
\ \
i {\cal M}^{\rm hard} ,
\label{EquationHadronicAmplitude}
\end{eqnarray}
in which the parton momenta are related by
Eq. (\ref{EquationRelationXXPrime}),
(\ref{EquationRelationYYPrime}) and
\begin{equation}
x_i = (1 - \xi_1 ) x'_i , \hspace{.25in}
y_i = (1 - \xi_2 ) y'_i , \hspace{.25in}
i=2 \cdots n.
\label{EquationRelationHigherXXPrimeYYPrime}
\end{equation}
These soft--matrix elements are precisely those found in our previous
work\cite{LuMilana} on dijet diffractive events at HERA.

As in \cite{LuMilana}, in order to continue we need to introduce approximations
and estimate the soft matrix elements contained in the
previous equation. Guided by the `handbag' appearance
of the amplitude in Fig. \ref{fig2}, we note in the
case of the helicity conserving amplitudes the resemblance
of the soft matrix elements to the gluon's structure function
\cite{Mueller,CarlsonMilana}:
\begin{equation}
G_{g/p} (x_1) = \sum_{n} \sum_{a} \int [dx] [d^2k_{\perp}] \mid
\psi_n(x,k_{\perp}) \mid^2 \delta (x_a - x_1).\label{struct}
\end{equation}
The second sum is over all gluons in a given Fock component.
For the case at hand that $\xi \ll 1$, we note from
Eq. (\ref{EquationRelationHigherXXPrimeYYPrime}) that
$x'_i \approx x_i$ for all $i = 2 \cdots n$,
and for $x'_1 > \xi_1$. And similarly
$y'_i \approx y_i$ for all $i = 2 \cdots n$,
and for $y'_1 > \xi_2$.
We also make the approximation $\tilde x_1 \approx x_1$
and $\tilde y_1 \approx y_1$.
For the helicity conserving processes,\footnote{As in \cite{LuMilana}, we
assume that the gluon structure function is quantitatively the largest
soft--matrix element so that helicity flipping events (in which the
polarizations of the gluons leaving and re--entering the proton are opposite)
are taken to be suppressed.}
we therefore estimate Eq. (\ref{EquationHadronicAmplitude}) by
\begin{equation}
i {\cal M}
=
\int_{\xi_1}^1
\frac{d x}{x} G_{g/p}(x)
\int_{\xi_2}^1
\frac{d y}{y} G_{g/p}(y)
\ \
i {\cal M}^{\rm hard} ,
\end{equation}
where in our approximation, the hard amplitude
is given by
\begin{equation}
i {\cal M}^{\rm hard}
=
-\frac{i}{2}
 \frac{g_{\rm w} \alpha_{\rm s}^2}{m_W}.
\end{equation}

\section{Higgs Production Cross Section}

The expression for the cross section of the $p+p \to p+p+H$
process is given by
\begin{equation}
d \sigma =
\frac{ |{\cal M}|^2 }{2^8 \pi^3 s}
\delta(\xi_1 \xi_2 s - m_H^2)
d \xi_1 d t_1 d \xi_2 d t_2 ,
\label{CrossSection}
\end{equation}
where $t_1=(p_1-p'_1)^2$, $t_2=(p_2-p'_2)^2$,
and $m_H^2$ is the mass of the Higgs particle.
In the calculation of the hard amplitudes we have
used the diffractive limit ($t_1, t_2 \to 0$).
For the integration of $t_1$ and $t_2$ away
from the diffractive limit we include in
Eq. (\ref{CrossSection}) the form factors
$e^{b t_1}$ and $e^{b t_2}$, and use
$b \sim 4$ GeV$^{-2}$ (see Ref.
\cite{LuMilana,DonnachieLandshoff}.)  In our approach these form factors
 represent a modeling of the allowable (i.e. ``intrinsic'') transverse momentum
dependence of the active partons.
After integrating out these variables, we obtain
for the total cross section
\begin{equation}
\sigma
=
\frac{1}{256 \pi^2}
\frac{\alpha_{\rm w} \alpha_{\rm s}^4}
     {s^2 m_W^2 b^2}
\int_{m_H^2 / s}^1
\frac{d \xi}{\xi}
\
\Biggl[
   \int_\xi^1
   \frac{d x}{x} G_{g/p}(x)
\Biggr]^2
\
\Biggl[
   \int_{m_H^2 / \xi s}^1
   \frac{d y}{y} G_{g/p}(y)
\Biggr]^2.\label{totalcross}
\end{equation}

In order to simplify the above integrals, we use the following
parametrization for
the gluon structure function:\footnote{In \cite{LuMilana} when we considered
the case that $G_{g/p}(x) \propto 1/x^{3/2}$, we mistakenly did not change
 normalizations of the gluon structure function when going from the case
$G_{g/p}(x) \propto 1/x$.  Using the normalization of
(\ref{gluonstruct}) effectively
reduces our rates for this case by one order of magnitude in
Fig. 2 of \cite{LuMilana}.}
\begin{equation}
G_{g/p}(x,Q^2=M_H^2) = \frac{c (1-x)^5}{x^{3/2}}\label{gluonstruct}.
\end{equation}
This low--$x$ singular form is in general agreement with both theoretical
expectations\cite{Gribov,Ralston} at these $Q^2$ values, and also the most
recent data from HERA
\cite{H1Collaboration,ZeusCollaboration}.
We take $c \sim 0.9$, to reasonably match with the latest CTEQ \cite{CTEQ}
parametrizations at these $x$ and $Q^2$ values.

Observe that with the low--$x$ dependence of (\ref{gluonstruct}) and
the fact that (\ref{totalcross}) depends essentially
on the squares
of the gluon structure functions of each of the colliding hadrons,
we obtain that our
total cross--section increases linearly with $s$.
  Such a continued growth of $\sigma$ of course violates general
unitarity bounds and must eventually be halted,
either from explicit higher twist
corrections entering the gluon structure function\cite{Gribov},
or/and from potentially novel low--$x$ effects
occurring in hadronic collisions\cite{mefirst}.
At the present energies being considered however,
the cross section is, as we will next see, exceptionally small.
Hence it is unlikely unitarity constraints play much
of a role and we expect all such corrections to be ignorable.

In Fig. \ref{fig4} we plot the cross section
at an upgraded Fermilab energy ($\sqrt{s} \sim 4$ TeV)
and for the LHC ($\sqrt{s} \sim 14$ TeV).
As a posthumous gesture, we also plot
what the cross section would have been for the
SSC ($\sqrt{s} \sim 40$ TeV).
We used to obtain these plots: $\alpha_{\rm w} \sim 1/30$,
$\alpha_{\rm s} \sim 0.1$, $m_W = 80.2$ GeV, and,
as stated earlier, $c=0.9$, $b=4$ GeV$^{-2}$.

\section{Conclusions}

We have estimated the cross section for the exclusive diffractive
production of an intermediate mass ($m_H < m_t$) Higgs particle
at LHC and Fermilab upgrade energies using perturbative QCD methods.
  Due to the color-singlet nature of the hard event, we expect
factorization to be applicable and the soft
matrix elements entering the cross--section to be directly related to those
entering exclusive, diffractive dijet events in
$e p$ collisions at HERA.  Until such data is forthcoming, we have approximated
these soft--matrix elements in terms of the gluon structure functions of the
colliding hadrons in order to obtain a working estimate of the expected rates.
We find that even at the most favorable
conditions of the LHC (with a planned luminosity ${\cal{L}} = 1.7 \times
10^{34} \, \rm(cm^{-2}-sec^{-1})$ \cite{PartData}),
the obtained cross section
is at least several orders of magnitude below what one might consider a working
level of usefulness.

A couple of factors might yet ameliorate somewhat these estimates.
First, past experience\cite{Mueller,C.Z.,Stoler} with exclusive processes in
perturbative QCD indicates that the
typical scale for evaluating the running coupling is generally significantly
less than that
of the hard scattering probe.  Even with a generous reduction factor of two
orders of
magnitude, so that in our formulas, $\alpha_s(Q^2=M_H^2) \rightarrow
\alpha_s(Q^2=(M_H/100)^2)$, this increases our
estimate by at most by a factor of $(2)^4 = 16$.
A second enhancement factor not unreasonable to expect
comes from the experience with inclusive
hard scattering collisions in which a
 $K-$factor type of correction enters proton--proton vs. lepton--proton
collisions.
Guessing a typical value for $K \approx 2$, which should now however enter in
the
{\it amplitude} (Fig. \ref{fig2}), this would then lead to another
factor of 4 or so increase in our rate.
We thus find that altogether somewhere between one and two orders of
magnitude enhancement above the rates depicted in Fig. \ref{fig4}
might not be unreasonable to expect.
Unfortunately, even with such a total enhancement factor added,
this still fails to bring up the magnitude of the cross section
to what could be called experimentally useful levels (for
the lightest Higgs masses in Fig. \ref{fig4} these
enhancements would correspond to a
possible total of just a few such events
per year at the LHC).  Thus, despite the
potential cleanness of the experimental signal,
exclusive diffractive production does not appear to
provide a viable channel for the detection
of a standard model Higgs particle.

While hard double--diffractive scattering does not thus appear to be
a feasible mechanism for probing new fundamental physics
(i.e. Higgs discovery) we still believe the mechanism
to be interesting in its own right for understanding
the parton structure of the ``pomeron'' and addressing
issues of factorization.  We thus emphasize the importance of
the experimental search for dijet double--diffractive events,
a theoretical study of which is now being pursued to estimate
expected yields at presently operating hadronic colliders,
as well as those anticipated in the not too distant future.

\noindent\underline{Acknowledgements}: We thank M. Banerjee, W. Broniowski,
and M. Sher for their input.  One of us (JM) would also like to thank
Rebecca Celia for her many spirited discussions.
This work was supported in part by the U.S. Department of Energy,
under grant No. DE--FG02--93ER--40762.

%%%%%%%%%%%%%%%%%%%%%%%%%%%%%%%%%%%%%%%%% References

\newpage
%%%%%%%%%%%%%%%%%%%%%%%%%%%%%%%%%%%%%%%%%
\begin{figure}
\caption{Effective gluon-Higgs coupling vertices induced
by a heavy top-quark loop.}
\label{fig1}
\end{figure}

\begin{figure}
\caption{Mechanism for hard-diffractive production of Higgs particle.
The dominant contribution comes from the subprocess where the Higgs
particle is produced through the hard diffraction of two parton-gluons.
$p_1$, $p_2$ are the initial proton momenta, and $p'_1$, $p'_2$
are the corresponding final momenta. }
\label{fig2}
\end{figure}

\begin{figure}
\caption{Feynman diagrams for the calculation of
the hard scattering amplitude, representing respectively
the (a) $s$-channel (7 diagrams), (b) $u$-channel (7 diagrams),
and (c) four-gluon-vertex (5 diagrams) exchange. There
is no $t$-channel color-singlet exchange. The Higgs line
can be attached to any gluon-line, three-gluon vertex,
or four-gluon vertex. The effective $ggH$, $gggH$ and $ggggH$
interaction vertices are given in the text.}
\label{fig3}
\end{figure}

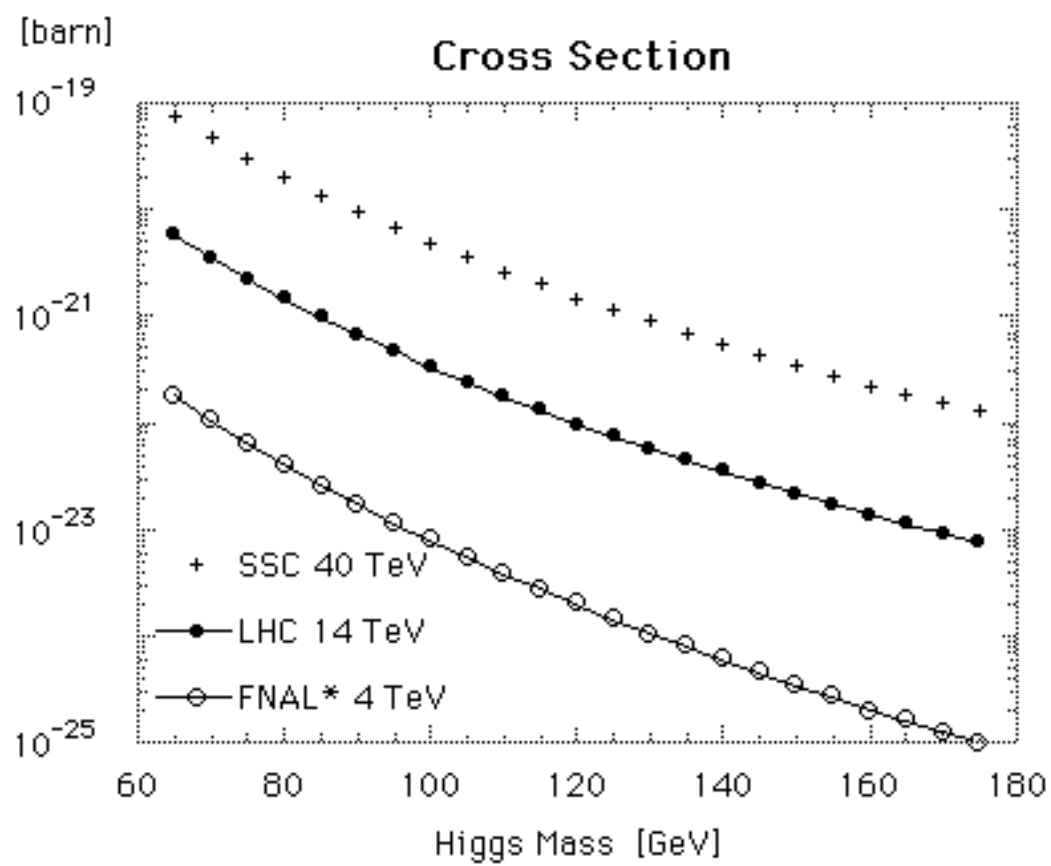
\begin{figure}
\caption{Cross section for the exclusive production of
Higgs at various hadron colliders (Fermilab upgrade and
LHC).   The SSC case has been included solely for the
melancholy.}
\label{fig4}
\end{figure}

\end{document}